\documentclass[twocolumn,showpacs,amsmath,amssymb,pre]{revtex4}
\usepackage{graphicx}

\begin{document}

\title{Thermal transistor: Heat flux switching and modulating}

\author{Wei Chung \textsc{Lo}$^{1,4}$,
Lei \textsc{Wang}$^{1,2}$\thanks{E-mail address:
phywangl@nus.edu.sg} and Baowen \textsc{Li}$^{1,3}$}

\affiliation{$^{1}$Department of Physics and Centre for
Computational Science and Engineering, National University of
Singapore,
Singapore 117542, Republic of Singapore\\
$^{2}$Department of Physics, Renmin University of China, Beijing 100872, P. R. China \\
$^{3}$NUS Graduate School for Integrative Science and Engineering,
Singapore 117597, Republic of Singapore \\
$^{4}$Institute of High Performance Computing, 1 Science Park
Road, Singapore 117528}

\begin{abstract}
 Thermal transistor is an efficient heat control device
which can act as a heat switch as well as a heat modulator. In
this paper, we study systematically one-dimensional and
two-dimensional thermal transistors. In particular, we show how to
improve significantly the efficiency of the one-dimensional
thermal transistor. The study is also extended to the design of
two-dimensional thermal transistor by coupling different
anharmonic lattices such as the Frenkel-Kontorova and the
Fermi-Pasta-Ulam lattices. Analogy between anharmonic lattices and
single-walled carbon nanotube is drawn and possible experimental
realization with multi-walled nanotube is suggested.
\end{abstract}


\maketitle

\section{Introduction}
Much attention has been devoted in the past years to the study of
heat conduction in low-dimensional systems \cite{Bonetto}. Such
study is not only important for understanding the fundamentals of
statistical mechanics, but also for potential applications in heat
control and management in nanoscale devices.

Indeed, various models for thermal rectifiers/diodes that allow
heat to flow easily in one direction have been proposed
\cite{rec,diode,Hu1,Hu2,nitzan,J,J2,yang}. Moreover, based on the
phenomenon of the \emph{negative differential thermal resistance}
($\emph{NDTR}$) observed in nonlinear lattices, a theoretical
model for ``thermal transistor" is also proposed \cite{trt}, which
allows us to control heat current (due to phonons) by adjusting
the temperature of the gate terminal (or called ``gate
temperature"). This is similar to the electronic transistor which
controls the electric current by adjusting the gate voltage. We
should point out that our ``thermal transistor" is different from
the recent ``heat transistor", which controls the heat flux
(caused by electrons) via adjusting the gate voltage
\cite{heattrst}. Most recently, based on the thermal transistor,
the different thermal logic gats have been also realized
theoretically\cite{WangLi07}, which means that the phonons, the
heat carrier, can also be used to carry information and processed
accordingly.

The purpose of the current paper are two folds. First, we shall
discuss how to improve significantly the switching efficiency of
1D thermal transistor; Second, we shall extend the study from 1D
model to 2D which is closer to laboratory fabricated materials.

\section{One-dimensional model}

The 1D thermal transistor model consists of three segments
$\emph{S}$, $\emph{D}$, and $\emph{G}$ and each segment is a
Frenkel-Kontorova (FK) lattice \cite{braun,pok}. The model
configuration, which is shown in Fig. 1(a), is similar to the one
proposed in Ref.\cite{trt} except for an additional interface
coupling $k_{intS}$. In each segment, we regard the particle that
is coupled to heat bath as the first particle and the interface
particle as the last particle. The total Hamiltonian of the 1D
model writes
\begin{equation}
\label{eq:H} H=H_{S}+H_{D}+H_{G}+H_{int},
\end{equation}and the Hamiltonian of each segment can be written as
$H_{W}$=$\sum_{i=1}^{N_{W}}\frac{p_{W,i}^{2}}{2m} +
\frac{k_{W}}{2}(x_{W,i}-x_{W,i-1})^{2}-\frac{V_{W}}{(2\pi)^{2}}\cos
2\pi x_{W,i}$, with $x_{W,i}$ and $p_{W,i}$ denote the
displacement from equilibrium position and the conjugate momentum
of the $i^{th}$ particle in segment $W$, where $W$ stands for
$\emph{S}$, $\emph{D}$ or $\emph{G}$. The parameters $k$ and
$\emph{V}$ are the harmonic spring constant and the on-site
potential of the FK lattice. We couple the last particle of
segment $\emph{S}$, $\emph{D}$, and $\emph{G}$ to particle
$\emph{O}$ via harmonic springs. Thus
$H_{int}$=$\frac{p_{O}^{2}}{2m}-\frac{V_{O}}{(2\pi)^{2}}\cos 2\pi
x_{O}+ \frac{k_{intS}}{2}(x_{N_{S}}-x_{O})^2+
\frac{k_{intD}}{2}(x_{N_{D}}-x_{O})^2+
\frac{k_{intG}}{2}(x_{N_{G}}-x_{O})^2$. Fixed boundaries are used,
i.e., $x_{W,0}$=0 and each segment consists of 10 particles.

\begin{figure}[tb]
\begin{center}
\includegraphics{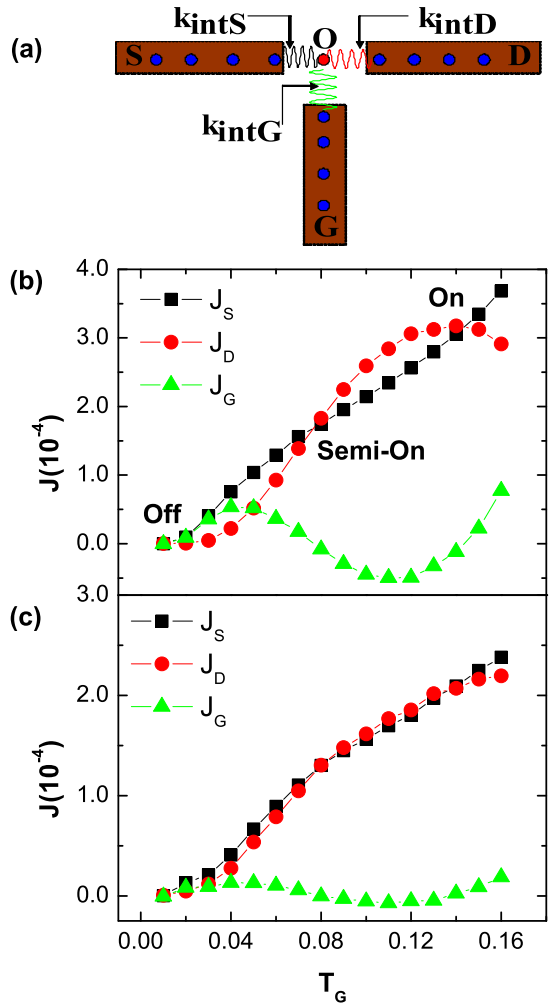}
\end{center}
\caption{(a) Configuration of the thermal transistor. Parameters
of the system are (b) Thermal switch: $T_{S}$=0.01, $V_{S}$=5.00,
$k_{S}$=0.35, $k_{intS}$=0.30; $T_{D}$=0.20, $V_{D}$=1.00,
$k_{D}$=0.20, $k_{intD}$=0.05; $V_{G}$=5.00, $k_{G}$=1.00,
$k_{intG}$=0.20, and $V_{O}$=5.00. (c) Thermal modulator: all
parameters remain the same as in (b) except with $k_{intG}$=0.05.}
\label{f1}
\end{figure}

In our simulations, we use the Nos\'{e}-Hoover heat baths
\cite{Nose} and we set the mass of each particle and the Boltzmann
constant to unity unless otherwise stated. In the nonequilibrium
stationary state, the heat flux, which is the rate of energy
transport, is constant along each segment and the local equilibrium
temperature is given by the time average of the square of particle's
velocity per degree of freedom.

We show in Fig. 1 that the model described by Hamiltonian (1) can
display different heat controlling features depends on the applied
gate temperature, $T_{G}$. In particular, the model can act as a
heat switch, namely, the system can be switched from a heat
insulator to a heat conductor by just adjusting the gate
temperature. As demonstrated in
Fig. 1(b), when $T_{G}$ is at 0.01, 0.08, and 0.14, 
$J_S$$=$$J_D$, thus $J_G$=0, namely, the control terminal does not
provide any current to the system, the system, however is switched
from an insulating (off) state, to a semi-conducting (semi-on)
state, and conducting (on) state. At these three points, $J_{D, S}$
are 3.8x$10^{-7}$, 1.8x$10^{-4}$, and 3.2x$10^{-4}$, respectively.
The ratio of $J_{D,S}$ at the ``on" state  to that at the ``off"
state is considered as the switching efficiency which is about 800.

We further demonstrate that the model can act as a heat modulator as
well in which $J_{D,S}$ can be continuously modulated by the gate
temperature. To show this, we recall that for any two weakly coupled
lattices, the heat current depends among others, on the strength of
the interface coupling used \cite{diode,PRB}. Thus, to have
$J_{D}$$\approx$$J_{S}$, we can reduce the strength of $k_{intG}$ so
that the heat flow through the control terminal decreases.
Apparently, $k_{intG}$ should not be too small otherwise $T_{O}$ is
not well-controlled by $T_{G}$. The numerical result is shown in
Fig. 1(c).

In our next configuration, we replace segment {\it D} with
a FPU-$\beta$ lattice whose phonon spectrum is contributed by low frequency
vibrations (acoustic phonons) only. The Hamiltonian is given by
$H$=$\sum_{i=1}^{N}\frac{p_{i}^{2}}{2m} +
\frac{k_{FPU}}{2}(x_{i}-x_{i-1})^{2} +
\frac{\beta}{4}(x_{i}-x_{i-1})^{4}$, with $\beta$ being the
anharmonicity. We show in Fig. 2(a) that such a setup reduces
the heat current at ``off" state greatly but affects the heat current at ``on" state
only slightly with the switching of about 1300, where $J_{D, S}$ equals to
2.15x$10^{-7}$ and 2.82x$10^{-4}$ at $T_{G}$=0.01 and 0.14,
respectively.  It can be noticed that for this and the former lattice choice in the present model, $J_{D,S}$ at the ``off" and ``on" state is about $10^{-7}$ and $10^{-4}$, respectively, with a switching of about $10^{3}$, this is a significant improvement
compared with previous one \cite{trt}. The high switching is mainly a result of the significant reduction of $J_{D,S}$ at the ``off" state,
which is basically due to the lower $T_O$ at ``off'' state ( in the previous work it is designed at 0.04).
Moreover, we found that using lattice with lower heat resistance as terminal {\it D} is ideal for achieving a large $J_{D}$.
In Fig. 2(b), we show the heat modulating effect with the same setup.
Since the demonstrated features of the transistor rely very much
on an efficient control of the interface temperature, it would be
interesting also to study how the property of the control terminal
can affect its functions. In Fig. 3, we show that when we use the
FPU-$\beta$ lattice as the control segment, due to the large heat
resistance in the segment, temperature $T_O$ does not well follow
$T_G$ thus heat currents in segment $\emph{S}$ and $\emph{D}$ do
not sensitively depend on $T_G$ in the low temperature regime.
Only when $T_{G}$$>$ 0.08, significant increment of currents and
continuous modulation of $J_{D,S}$ are observed. In other words,
$J_{D,S}$ can be even more precisely adjusted.

\begin{figure}[tb]
\begin{center}
\includegraphics{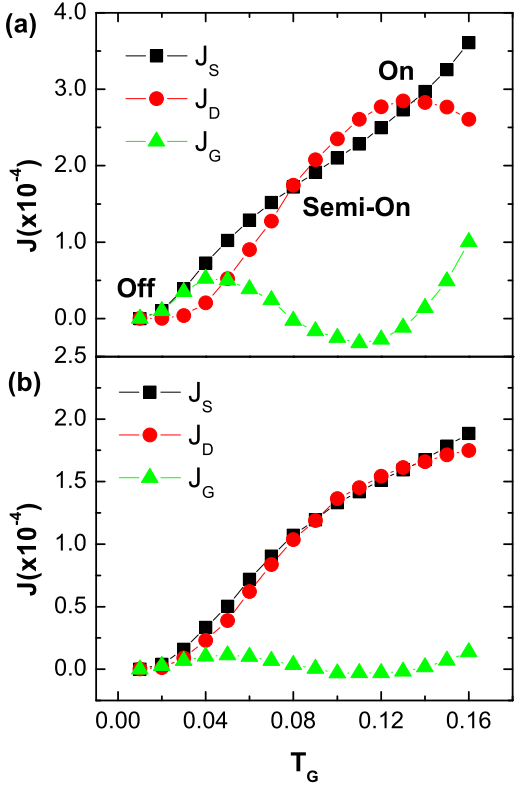}
\end{center}
\caption{(a) Thermal switch: a FPU-$\beta$ lattice with
$k_{D}$(=$k_{FPU}$)=0.20 and $\beta$=0.05 is used as segment
\emph{D}. Other parameters remain the same as in Fig. 1 (b). (b)
Thermal modulator: all parameters remain the same as in 2 (a)
except with $k_{intS}$=0.25, $k_{intD}$=0.055, and
$k_{intG}$=0.04.} \label{f2}
\end{figure}

\begin{figure}[tb]
\begin{center}
\includegraphics{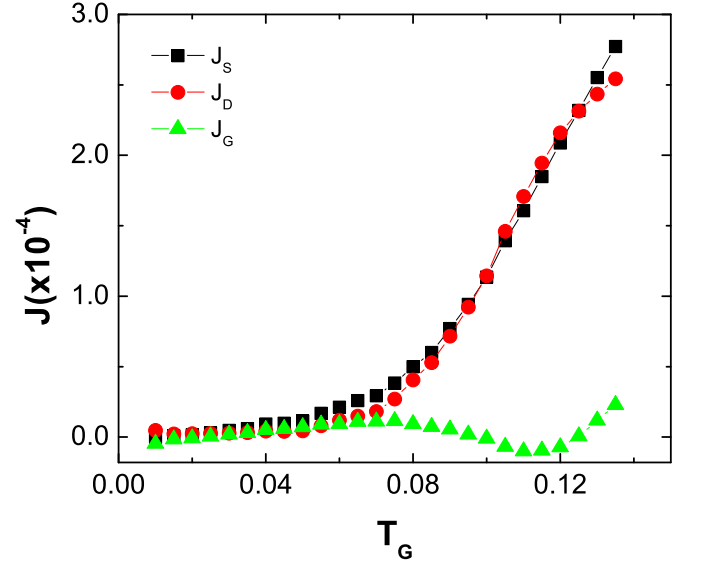}
\end{center}
\caption{The thermal modulator in which the FK lattice is used as
segment $S$ and the FPU-$\beta$ lattices are used as segment $D$
and $G$. The system parameters are $T_{S}$=0.01, $V_{S}$=5.00,
$k_{S}$=0.35, $k_{intS}$=0.60; $T_{D}$=0.20, $k_{D-FPU}$=0.20,
$\beta_{D}$=0.05, $k_{intD}$=0.05; $k_{G-FPU}$=0.20,
$\beta_{G}$=2.00, $k_{intG}$=0.20, and $V_{O}$=5.00.} \label{f3}
\end{figure}
The results presented so far have shown that by choosing the
lattice in different segment appropriately, one can improve the
switching and modulating efficiency significantly. This message is
very important and useful for optimization and experimental
realization of the thermal transistor. In the following, we shall
extend the study of the thermal transistor to 2D. Such extension
is important in application, as it takes into account the
interaction between longitudinal and transverse motions and
therefore is closer to real material such as thin film {\it etc}.
However, it is not our intention here to look for optimum
switching/ modulating condition for this model.

\section{Two-dimensional model}

The displacement from equilibrium position and the momentum of a
particle in 2D lattice are labeled as $\vec{q} \equiv \{q^{x},
q^{y}\}$ and $\vec{p} \equiv \{p^{x},p^{y}\}$, respectively. The
relative separation between two particles is $\Delta
r$=$|\vec{r}|$=$|\vec{q_1}-\vec{q_2}|$. The Hamiltonian of our model
consists of the Hamiltonian of each segment:
\begin{eqnarray}
\label{eq:2D1}
H_{W}=\sum_{i=1}^{N_{W,X}}\sum_{j=1}^{N_{Y}}[\frac{|\vec{p}_{{W},i,j}|^{2}}{2m}+
V_{W}(\Delta r_{W,i,j;i-1,j})\nonumber\\
+ V_{W}(\Delta r_{W,i,j;i,j-1})
- U_{W}(\vec q_{W,i,j})],\end{eqnarray}
and the Hamiltonian of the interface:
\begin{eqnarray}\label{eq:2D2}
H_{int}=\sum_{j=1}^{N_{Y}}[\frac{|\vec{p}_{O,j}|^{2}}{2m}-U_O(\vec
q_{O,j})
+V_{intS}(\Delta r_{N_{S,X},j;O,j})  \nonumber\\
+V_{intD}(\Delta r_{N_{D,X},j;O,j}) +V_{intG}(\Delta
r_{N_{G,X},j;O,j})].
\end{eqnarray}

Similarly, all segments are coupled to each other via harmonic
springs. In our model, segments $S$ and $G$ are FK lattices and
segment $D$ is FPU-$\beta$ lattice. Thus, $V_{S/G/intW}$($\Delta
r$)=$\frac{k}{2}$$\Delta r^{2}$, $V_{D}(\Delta
r)$=$\frac{k}{2}\Delta r^{2}$+$ \frac{\beta}{4}\Delta r^{4}$,
$U_{D}$=0, and $U_{S/G/O}(\vec q)$=$\frac{V_{S/G/O}}{(2\pi)^2}\cos
2\pi q^{x}\cos 2\pi q^{y}$. The harmonic spring constant $k$ and
the on-site potential $V$ may take different values in different
segments. Periodic and fixed boundaries are used in the Y
(transverse) and X (longitudinal) direction respectively, namely,
$\vec{q}_{i_{W},0}$=$\vec{q}_{i_{W},N_{Y}}$ and
$\vec{q}_{0_{W},j}$=(0, $j$) for $j$=$\{1,...,N_{Y}\}$. Thus the
topologies of the segments, as shown in Fig. 4(a), are cylinders.
Nos\'{e}-Hoover heat baths are coupled to the ends of segment
($i$=1). In the stationary state, temperature gradients are set up
in the $X$ direction and thus only heat currents through
longitudinal links, $J_{i_{W}}$(=$\sum_{j=1}^{N_{Y}}J_{i_{W},j}$)
need to be taken into account, which are constant along each
segment.

\begin{figure}[tb]
\begin{center}
\includegraphics{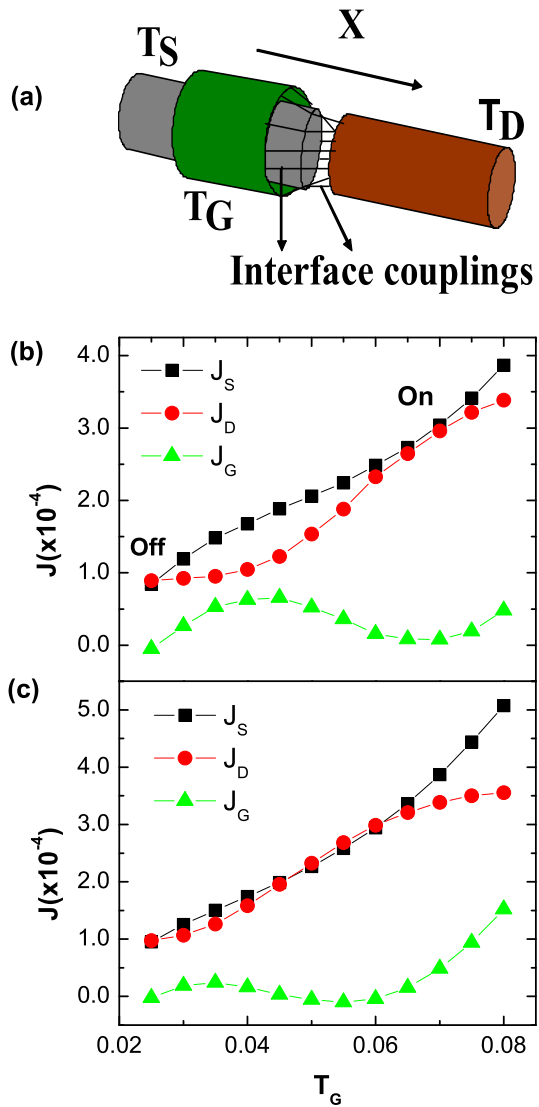}
\end{center}
\caption{(a) The prototype of a two-dimensional thermal
transistor. Parameters used are (b) Thermal switch: $T_{S}$=0.01,
$k_{S}$=0.20, $V_{S}$=5.00, $k_{intS}$=0.50; $T_{D}$=0.17,
$k_{D-FPU}$=1.00, $\beta$=3.00, $k_{intD}$=0.035;
$k_{G-FPU}$=1.00, $V_{G}$=6.00, $k_{intG}$=2.50, and $V_{O}$=5.00.
(c) Thermal modulator with $T_{S}$=0.015. Here,
$N_{D,X}$=$N_{G,X}$=10, $N_{S,X}$=9, and $N_{Y}$=4.} \label{f4}
\end{figure}

In Fig. 4(b), we show the heat switching for the 2D thermal
transistor model. It can be seen that when $T_{G}$ is at 0.01 and
0.07, $J_{S}$$=$$J_{D}$ and $J_{G}$=0. At these two points,
$J_{D,S}$ are 8.43x$10^{-5}$ and 3.38x$10^{-4}$ respectively and
thus the switching efficiency is about 4. This is a significant
reduction when compared with the 1D case and this scenario is also
quite similar to the 2D thermal rectifier \cite{J2}. We show in Fig.
4(c) that when $T_{S}$ is increased, the difference between $J_{S}$
and $J_{D}$ is reduced, which gives a better modulation of $J_{D,S}$
by the gate temperature. It is worth mentioning that in 2D case,
the crossover energy, particles above which are able to overcome the on-site potential barrier is
much lower than that in 1D case. Thus the lattice parameters should be re-designed,
otherwise the switching efficiency is reduced significantly as shown above.

\section{Discussion and Conclusion}

In summary, we have provided several ways in improving the
switching efficiency in 1D thermal transistor, namely, one can
change different lattice in different segments, and/or by
adjusting the coupling strength at the interface. Our study has
been extended to two-dimensional lattices and a prototype of 2D
thermal transistor - a tube-like model have been proposed. Our
numerical simulations have demonstrated that such a prototype can
work as a heat switch and a heat modulator - two basic functions
of a thermal transistor.

In our numerical simulations, we have used dimensionless units for
the sake of computations. However, all dimensionless units are
related to the dimensional physical variables through dimensional
scaling and consequently, the physical (real) temperature $T_r$ is
related to the numerical temperature {\it T} through $T_r=\frac{m
\omega_{o}^{2}a^{2}}{k_{B}}{T}$ where $k_{B}$ is the Boltzmann
constant (please refer to \cite{Bambi1} for details). Suppose we
use single-walled carbon nanotube as our experimental prototype
for the thermal transistor model, namely with the mass of particle
$m$, lattice constant $a$, and natural oscillation frequency
$\omega_{o}$ correspond to the real values of such a material
\cite{parameters}, ${T}$=0.01 in our simulations corresponds to
physical temperature $T_r$ of about $100$ K. Similar analysis for
heat flux gives $J_r$ (=$ m \omega_o^{3} a^{2} {J}$)
$\thicksim$$10^{-5}$${J}$. Therefore, with ${J}$
$\thicksim$$10^{-4}$, $J_r$ is about $10^{-9}$ W (J/s) which is
smaller than that one in Ref.\cite{Zhang}. The smaller heat flux
obtained in our model is not surprising. It is mainly due to the
interacting with external on-site potential \cite{D} and the
interface thermal resistance arises in coupling dissimilar
nonlinear lattices \cite{J}, which is similar to the case of
electronic diode where the electrical current are greatly reduced
by the P-N junction. For a homogeneous FPU lattice with similar
parameters, $J_r$ can be about 2 orders of magnitude higher.

Even though a real working prototype for solid state thermal
transistor is still an open issue, we believe that the above model
or its variants will be realized sooner or later. Moreover, a solid
state thermal rectifier using nanotubes has been experimentally
demonstrated \cite{chang1} (see also Ref\cite{wuli} for other
variants). More recently, a tunable and reversible thermal link
using multi-walled carbon nanotubes (MWCNTs) \cite{chang2} were
successfully demonstrated whereby the thermal conductance can be
controlled mechanically by displacing the outer shell with respect
to the inner one. This is a step forward to the realization of
thermal transistor. In fact, in the heat control models, be it a
thermal rectifier or a thermal transistor, the key factors are: (1)
the broken of the spatial symmetry of terminal, namely they are two
different materials; (2) nonlinearity in each segment. Therefore,
the possible material to be used for the thermal transistor might be
multiwalled nanotubes. As the recent experiment \cite{chang2} shows
that the outer wall can be used to control the thermal conductance
of the nanotube, thus it can be used as the controlling terminal G.

\section*{Acknowledgment}
The work is supported partially by an academic research fund,
R-144-000-203-112 from MOE of Singapore, and the Temasek Young
Investigator Award of DSTA Singapore under Project Agreement
POD0410553

\end{document}